\documentclass[aps,prl,superscriptaddress,11pt]{revtex4}

\usepackage[T1]{fontenc}
\usepackage{times,verbatim}
\usepackage{color,graphicx}
\usepackage{subfig,array}
\usepackage[latin1]{inputenc}
\usepackage[driverfallback=dvipdfm,colorlinks=true,linkcolor=blue,anchorcolor=blue,citecolor=red,urlcolor=magenta]{hyperref}
\usepackage{amsthm,amssymb,amsmath}

\newtheorem{thm}{Theorem}

\newtheorem{proposition}[thm]{Proposition}

\newtheorem{prop}[thm]{Proposition}

\newtheorem{lemma}[thm]{Lemma}

\newtheorem{rem-eg}[thm]{Remark and Example}

\newcommand{\K}{\mathcal{K}}

\def\IC{{\mathbb C}}
\def\IR{{\mathbb R}}
\def\IN{{\mathbb N}}

\def\cE{{\cal E}}

\def\bc{{\bf c}}
\def\bd{{\bf d}}

\def\b0{{\bf 0}}
\def\bx{{\bf x}}

\def\cB{{\cal B}}

\def\cD{{\cal D}}
\def\cE{{\cal E}}

\def\cB{{\cal B}}

\def\tr{{\rm tr}\,}

\def\diag{{\rm diag}}

\def\[{\left [}
\def\]{\right ]}
\def\({\left (}
\def\){\right )}

\def\<{{\langle}}
\def\>{{\rangle}}
\def\1{{\bf 1}}
\def\bc{{\bf c}}
\def\bx{{\bf x}}

\newcommand{\cI}{\mathcal{I}}
\newcommand{\cJ}{\mathcal{J}}

\def\qed{\hfill\vbox{\hrule width 6 pt \hbox{\vrule height 6 pt width 6 pt}}\medskip} 

\begin{document}


\title{Coherence measures induced by norm functions}

\author{Yangping Jing}
\affiliation{School of Science,
Hangzhou Dianzi University,
Hangzhou, 310018,  Zhejiang, P.R. China.
jyp@hdu.edu.cn}
\author{Chi-Kwong Li}
\affiliation{Department of Mathematics, College of William and Mary,  Williamsburg, VA, USA  23187. ckli@math.wm.edu}
\author{Edward Poon\footnote{Corresponding author}}
\affiliation{Department of Mathematics, Embry-Riddle Aeronautical
University, Prescott, AZ 86301, USA.
poon3de@erau.edu}
\author{Chengyang Zhang}
\affiliation{
School of Mathematics and Information Science,
Shaanxi Normal University, Xi'an, 710119,
P.R. China.
zhangcy@snnu.edu.cn}

\date{Revised version submitted March 9, 2021}

\begin{abstract}
Which matrix norms induce proper measures for quantifying quantum coherence?  We study this problem for two important classes of norms and show that (i) coherence measures cannot be induced by any unitary similarity invariant norm, and (ii) the $\ell_{q,p}$-norm induces a coherence measure if and only if $q=1$ and $1 \leq p \leq 2$, thus giving a new class of coherence measures with simple closed forms that are easy to compute.  These results extend and unify previously known facts about norm-induced coherence measures, and lead to a broader framework for understanding what functionals can be coherence measures. 
\end{abstract}
\maketitle

\section{1. Introduction}

Due to its intimate connection to the superposition principle, quantum coherence 
plays a central role in quantum mechanics---indeed, it is well known that quantum mechanical systems differ significantly from classical systems mainly because of coherence.  As such, coherence is an important physical resource in quantum information and quantum computation \cite{NC}. It also plays an important role in a wide variety of other research fields, such as quantum biology \cite{2,3,4,5,6,7}, nanoscale physics \cite{8,9}, and quantum metrology \cite{10,11}.

One of the most widely used frameworks for quantifying coherence was introduced in \cite{Bau} as four key conditions; a useful equivalent formulation is given in \cite{Yu}.  We shall use these frameworks as our starting point, leaving the details for the motivation and background for these frameworks to the aforementioned references.

To state the conditions in \cite{Bau} we shall represent quantum states by density matrices.  It is important to note that coherence is basis dependent---the reference basis with respect to which coherence is measured depends on the physical problem.  We henceforth fix a basis (which must be compatible across all dimensions) so that the incoherent states are the diagonal density matrices.  Let $M_{m,n}$ be the linear space of $m \times n$ complex matrices (and write $M_n$ if $m=n$).  Let $\cD_n$ be the set of density matrices in $M_n$; let $\cI_n$ be the subset of diagonal density matrices, so the set of incoherent states in $M_n$ is precisely $\cI_n$.  States not in $\cI_n$ are said to be coherent states.
 
One of the conditions in both frameworks requires monotonicity of the coherence measure under certain quantum operations.  In general, a quantum operation transforming quantum states in $M_n$ to quantum states in $M_m$ is a trace preserving completely
positive map $\mathcal{E}: M_n \rightarrow M_m$ admitting the
following operator sum representation:
$$\cE(A) = \sum_{j=1}^r K_j A K_j^{\dag}  \quad \hbox{ for all } A \in M_n.$$
Here $r$ is a positive integer and $K_1, \dots, K_r\in M_{m.n}$ (called the Kraus operators corresponding to $\cE$)
satisfy $\sum_{j=1}^r K_j^{\dag}K_j=\textmd{I}_n$ (the $n \times n$ identity matrix); one may see \cite{NC} for the
general background of Kraus operators.  If $K_j\rho {K_j}^\dag$ is a diagonal matrix for all indices $j$ and all 
$\rho \in \cI_n$ then $\cE$ is an incoherent operation and $\{K_1,\dots, K_r\}$ is a set of incoherent Kraus operators.
 
We can now state the four defining conditions in \cite{Bau} which a real-valued function $C$ must satisfy to be a coherence measure.
\begin{itemize}
\item[(B1)] If $\rho \in \cD_n$, then
$C(\rho)\geq 0$; the equality $C(\rho)=0$ holds
if and only if $\rho \in \cI_n$.
\item[ (B2)]
If $\Lambda: M_n \rightarrow M_m$
is an incoherent operation and $\rho \in \cD_n$,
then $C(\rho)\geq C(\Lambda(\rho))$.
\item[(B3)]
Suppose $\rho \in \cD_n$ and $\Lambda: M_n \rightarrow M_m$ is an incoherent operation
with incoherent Kraus operators $K_1, \dots, K_r$.
If
$p_j=\tr(K_j\rho K_j^\dag)$ and
$\rho_j=\frac{1}{p_j} K_j\rho K_j^\dag$
for $j = 1, \dots, r$,  then
$C(\rho)\geq \sum_{j=1}^r p_jC(\rho_j).$
\item[(B4)] For any $\{\rho_1, \dots, \rho_r\} \subseteq \cD_n$ and
 any probability distribution $\{p_1, \dots, p_r\}$,
 $\sum_{j=1}^r p_jC(\rho_j)\geq C(\sum_{j=1}^r p_j\rho_j).$
\end{itemize}
Because incoherent operations may transform states on an $n$-dimensional space into states on an $m$-dimensional space, $C$ must be defined on density matrices of any size; this explains our earlier remark that the fixed incoherent basis must be compatible across all dimensions.

Note that condition (B3) combined with (B4)
automatically imply (B2). In general,
a real-valued function $C$ is called a coherence measure if it satisfies the above four conditions; if only the conditions (B1), (B2), and (B3) are satisfied, $C$ is usually
called a coherence monotone.  Researchers have considered different kinds of coherence measures; see \cite{Bau,SXF,Streltsov,GD,DB,Yu,Piani,
Napoli,RAE,AD,EG,XZCM,Rana,BUF,Cs,QG,HF}. 

Because a state is incoherent if and only if it is diagonal, a very natural coherence measure is one which measures the distance between a given state and the set of diagonal states.  For example, suppose $\nu$ is a norm defined on matrices of arbitrary sizes.  Given $\rho \in \cD_n$, define
\begin{equation}
\label{C-nu}
C_{\nu}(\rho) = \min\{ \nu(\rho-\sigma): \sigma\in \cI_n\}.
\end{equation}
It is not hard to see that $C_{\nu}$ automatically satisfies (B1) and (B4), and (B2) will hold if the norm $\nu$ is
contractive under any incoherent map $\Lambda$, i.e.,
$\nu(\rho_1- \rho_2) \ge \nu(\Lambda(\rho_1)-\Lambda(\rho_2))$; one may find the details in \cite{Bau}.  However, condition (B3) is more difficult to determine, and its status depends on the particular norm.

In \cite{Bau}, it was shown that if $\nu$ is the $\ell_1$-norm
then $C_{\nu}(\rho) = \sum_{i\ne j} |\rho_{ij}|$ is a
coherence measure. In \cite{Rana}, the
authors showed that $C_\nu$ is not a  coherence measure
if $\nu$ is the
$\ell_p$-norm $\nu(A) = (\sum_{i,j}|a_{ij}|^p)^{1/p}$ for $p > 1$.
In the same paper, the authors also showed that
$C_\nu$ is not a coherence measure if $\nu$ is  the
Schatten $p$-norm $\nu(A) = (\tr|A|^p)^{1/p}$ with $p > 1$,
where $|A|$ is the positive semi-definite matrix $A$ such that
$|A|^2 = A^\dag A$.
When $p = 1$, the Schatten $p$-norm reduces to the trace norm
$\|A\| = \tr |A|$, which is used frequently in quantum
information science research.
At one point, researchers believed  that $C_\nu$ is a coherence measure if $\nu$ is the trace norm \cite{CGJLP,Rana}.  However, this is not the case, as shown in \cite{Yu} by using the following alternative (but equivalent) framework for coherence measures.

\begin{itemize}
\item[(C1)] If $\rho \in \cD_n$, then
$C(\rho)\geq 0$; the
equality  $C(\rho)=0$ holds if and only if $\rho\in \cI_n$.
\item[(C2)] If $\Lambda: M_n \rightarrow M_m$ and $\rho \in \cD_n$, then
$C(\rho)\geq C(\Lambda(\rho))$.
\item[(C3)]
Let $\rho_1 \in \cD_{n_1}, \rho_2\in \cD_{n_2}$,
$p_1, p_2 \ge 0$ satisfy $p_1+p_2=1$.
Then
$C(p_1 \rho_1 \oplus p_2 \rho_2)=p_1C(\rho_1)+p_2C(\rho_2)$.
\end{itemize}

Clearly, (B1)-(B2) and (C1)-(C2) are the same.
The advantage of this alternative framework is it
changes the two inequalities in (B3)-(B4)
into a concrete equality (C3). In addition to showing that
the trace norm does not induce a  coherence measure,
it was shown in \cite{Yu} that
one can modify the trace norm to define a coherence measure
using the definition
$$C(\rho) = \min\{ \|\rho-t\sigma\|_1: t \ge 0, \ \sigma \in \cI_n\}
\quad \hbox{ for } \rho \in \cD_n.$$
However, it was shown in \cite{JLP} that this coherence measure has some limitations.

We shall use both sets of conditions 
(B1)-(B4) and (C1)-(C3) to study coherence measures as they provide different advantages in different situations.
For example, if one knows (or assumes) that a function is a coherence measure, one then has the powerful equality 
(C3); conversely, to show that a function is a coherence measure it is often easier to prove the inequalities in 
(B3)-(B4). 

\medskip

In this paper, we provide a more general view on 
coherence measures induced by norms. Instead of
studying a specific norm, we consider a class of norms
which will induce (or not induce) proper coherence measures.
The study will deepen the understanding of the fundamental
properties of a distance measure between a state and 
the set of classical states that will yield a coherence
measure. In particular,
we extend the results for the trace norm and the $\ell_1$-norm, obtaining two general results on coherence measures induced by norm functions.  Note that the trace norm is an example of a unitary similarity invariant (USI) norm (that is, a norm $\nu$ satisfying $\nu(A) = \nu(U^\dag AU)$ for any $A \in M_n$ and unitary $U \in M_n$).  In fact, all Schatten $p$-norms are USI norms.

In Section 2, we show that if $\nu$ is a USI norm, then 
$C_\nu$, given in (\ref{C-nu}),
cannot be a coherence measure.  As an immediate consequence, we recover the fact that a Schatten $p$-norm does not
induce a coherence measure for any $p \in [1,\infty]$.  However this result rules out many, many more potential candidates---there is a huge variety of USI norms, including the $\ell_2$-norm and Ky Fan $k$-norms.  In particular, this shows the impossibility of inducing a coherence measure from a norm that is invariant under time evolution.

In Section 3, we generalize in a different direction, by considering the $\ell_{q,p}$-norm on $M_n$, defined by
$$\ell_{q,p}(A) = \ell_q(\ell_p(A_1), \dots, \ell_p(A_n))$$
for $p,q \in [1,\infty]\times [1,\infty]$,
where $A_1, \dots, A_n$ are the columns of a matrix $A \in M_n$; recall that for $\bx = (x_1, \dots, x_n)$,
$$\ell_\infty(\bx) = \max\{|x_j|: 1 \le j \le n\},
\quad \hbox{ 
and } \quad
 \ell_r(\bx) = (\sum_{j=1}^n |x_j|^r)^{1/r}
\quad \hbox{ for } \ r \in [1, \infty).$$
Note that the $\ell_p$-norm is precisely the $\ell_{p,p}$-norm.  One may see \cite{KL} for some general background for the $\ell_{q,p}$-norm.

For simplicity, write $C_{q,p}$ for $C_{\ell_{q,p}}$; that is, 
$$C_{q,p}(\rho) = \min\{ \ell_{q,p}(\rho - \sigma): \sigma \in \cI_n\}.$$
We will show that $C_{q,p}$ is a coherence measure
if and only if $q = 1$ and $p \in [1,2]$.
As an immediate consequence of our result, we recover the fact that the $\ell_p$-norm induces a coherence measure if and only if $p = 1$.  Our result also exhibits a new class of coherence measures that are easy to compute and match one's intuition of depending only on the off-diagonal entries.  For example, to compute $C_{1,2}(\rho)$, replace the diagonal entries of $\rho$ by zero and add up the Euclidean norms of the columns of the resulting matrix.  Note that $C_{1,2}$ also has better smoothness properties than the important measure $C_{\ell_1}$ and, given that $p=2$ forms the boundary between norms which induce coherence measures and norms that don't, $C_{1,2}$ may be a sharper measure than $C_{\ell_1}$.

One might wonder why some $\ell_{q,p}$-norms induce coherence measures while no USI norm does.  Some physical quantities, such as von Neumann entropy, are unaffected by unitary similarities.  However, since coherence
measures represent the "distance" between a quantum state and the set of classical (diagonal) states, and since
one can always find a unitary similarity transformation converting a given quantum state into a diagonal state, it is
not surprising that unitary similarity transformations can change this "distance"; the result on USI
norms confirms the fact that there are no good connections between coherence measures and unitary similarity.

\section{2. Coherence measures and unitary similarity invariant norms}

Recall that a norm on $M_n$ is unitary similarity invariant (USI) if $\|U^\dag AU\| = \|A\|$ for any $A \in M_n$ and unitary $U \in M_n$.
Clearly, if $A$ is Hermitian, then $\|A\|$ depends only on the eigenvalues of $A$.
We will need the following important fact 
(see \cite[Theorem 3.3]{LT2} and also \cite[Theorem 4.1]{LT} for more details): if $\|\cdot\|$ is a USI norm on the space
of $n\times n$ Hermitian
matrices, then there is a compact set $S_n \subseteq \IR^{1\times n}$,
depending on the norm $\|\cdot\|$,
such that the following two conditions hold.
\begin{itemize}
\item[
(N1)]  If $(c_1, \dots, c_n) \in S_n$ then
$\pm (c_1, \dots, c_n)P \in S_n$ for any permutation matrix $P \in M_n$.
\item[
(N2)]  For every Hermitian matrix $A$ with eigenvalues
$\lambda_1(A) \ge \cdots \ge \lambda_n(A)$,
there is a vector $(c_1, \dots, c_n) \in S_n$
with $c_1 \ge \cdots \ge c_n$ such that
$$\|A\| =
\sum_{j=1}^n c_j \lambda_j(A)
= \max\left\{ \sum_{j=1}^n d_j \lambda_j(A):
(d_1, \dots, d_n) \in S_n  \right\}.
$$
\end{itemize}

\medskip

The Schatten $p$-norms are USI norms; another example is the numerical radius, defined by
$$w(A) = \max\{ |\tr A \rho|: \rho \in \cD_n\} \qquad \hbox{ for }
\quad A \in M_n.$$
It is interesting to note that the set $S_n$ satisfying (N1)-(N2)
for the numerical radius can be
$\{\pm (1,0,\dots, 0)P: P \in M_n \hbox{ is a permutation matrix}\}$;
the set $S_n$ satisfying (N1)-(N2) for the Schatten $p$-norm
can be
$\{(d_1,\dots, d_n)\in \IR^{1\times n}:
\sum_{j=1} |d_j|^q = 1\}$ with $q = (1-1/p)^{-1}$ if $p > 1$,
and
$\{(d_1, \dots, d_n)\in \IR^{1\times n}: \max_j |d_j| = 1\}$
if $p = 1$.

As mentioned in the introduction,
if $\nu$ is the Schatten $p$-norm with $p \ge 1$,
then $C_\nu$ is not a coherence measure.
We will prove the following much more general result.

\begin{thm}  \label{2.1}
There is no  coherence measure $C_\nu$ induced by a USI
norm $\nu$.
\end{thm}

To prove this theorem we begin by establishing two auxiliary propositions which are of independent interest and will be used in Section 3.

\begin{proposition} \label{2.2}
Suppose $\|\cdot\|$ is a norm on $M_n$.
Define $C: \cD_n\rightarrow [0, \infty)$ by
$$C(\rho) = \min \{ \|\rho - \sigma\|: \sigma \in \cI_n\}.$$
Then $(B1)$ and $(B4)$ hold. If $(B3)$ holds, then so does $(B2)$.
\begin{itemize}
\item[{\rm (a)}] Assume $(B2)$ holds. If $P \in M_n$ is a permutation
or a diagonal unitary matrix, then
$C(\rho) = C(P^\dag \rho P)$.
\item [{\rm (b)}] If   $\|\cdot\|$ is an absolute norm, i.e.,
$\|(a_{ij})\| = \|(|a_{ij}|)\|$, then
$C(\rho)  = \|\rho-\rho_\diag\|$.
\end{itemize}
\end{proposition}

\noindent
\it Proof. \rm As noted in Section 1, basic norm properties imply conditions (B1) and (B4).  If (B3) holds, then (B3) and (B4) imply that (B2) holds.

(a) If (B2)  holds, then we may consider the incoherent
operation $\Lambda(A) = P^\dag AP$ for a permutation
or a diagonal unitary matrix $P$. Then $C(\rho) \ge C(P^\dag \rho P)$.
Now, consider $\tilde \rho = P^\dag \rho P$ and
$\tilde \Lambda(A) = PAP^\dag $.
Then
$$C(P^\dag\rho P) =
C(\tilde \rho) \ge C(P\tilde \rho P^\dag) = C(\rho).$$

(b) Suppose  $\|\cdot\|$ is an absolute norm.
Let $\sigma^*  \in \cI_n$ satisfy
 $\|\rho-\sigma^*\| = \min\{ \|\rho-\sigma\|: \sigma \in \cI_n\}$.
 Replace the diagonal entries of $\rho - \sigma^* $ by their negative
 to obtain $\tau$. Then
 $$\|\rho - \sigma^*\| = \|\tau\|$$ and
$$
\|\rho-\sigma^*\| \le \|\rho - \rho_\diag\|
 = \frac{1}{2} \|(\rho-\sigma^* )+\tau\| 
  \le \frac{1}{2} (\|\rho-\sigma^*\| + \|\tau\|)
 = \|\rho-\sigma^*\|.
$$
\vskip -.38in \qed

\begin{proposition} \label{2.3}
Suppose $\|\cdot\|$ is a norm on $M_n$ such that
$\|P^\dag A P\| = \|A\|$ for any trace zero
Hermitian matrix $A\in M_n$
and permutation matrix $P \in M_n$. Let
$$C(\rho) = \min\{ \|\rho - \sigma\|:
\sigma \in \cI_n\}.$$
Suppose $R_\ell = E_{12} + \cdots + E_{\ell,1} \in M_\ell$
is the basic circulant.
If $\rho = \rho_1 \oplus\cdots\oplus \rho_k \in \cD_{n_1}
\oplus \cdots \oplus \cD_{n_k}$ satisfies
$R^\dag \rho R = \rho$ with $R = R_{n_1} \oplus \cdots \oplus R_{n_k}$,
then
$$C(\rho) = \|(\rho_1 - s_1I_{n_1}) \oplus \cdots \oplus
(\rho_{k} - s_k I_{n_k})\|$$
for some $s_1, \dots, s_k \in \IR$.
If $\rho_p = \rho_q$, we may assume that $s_p = s_q$.
In particular, if $\rho_1 = \cdots = \rho_k$, then
$C(\rho) = \|\rho-\rho_\diag\|$.
\end{proposition}

\noindent
 \it Proof. \rm  Suppose $\|\cdot\|$ 
and $\rho= \rho_1\oplus \cdots \oplus \rho_k\in \cD_n$
 satisfy the hypotheses of the proposition.
Let $\sigma^*  \in \cI_n$ satisfy
 $\|\rho-\sigma^*\| = \min\{ \|\rho-\sigma\|: \sigma \in \cI_n\}$.
Since $R^\dag \rho R = \rho,$
we see that every $\rho_j$ has constant diagonal entries.
Let $N = {\rm lcm}(n_1, \dots, n_k)$, the least common multiple of
$n_1,\dots, n_k$.
If $\tilde \sigma = \frac{1}{N}\sum_{j=1}^N (R^j)^\dag\sigma^* R_j$, then
$$
\| \rho-\sigma^* \|
\le \|\rho-\tilde \sigma\|
= \frac{1}{N} \| \sum_{j=1}^N  (R^j)^\dag (\rho - \sigma^*)R^j\|
\le \frac{1}{N} \sum_{j=1}^N \|(R^j)^\dag (\rho - \sigma^*)R^j\|
= \|\rho-\sigma^*\|.
$$
Thus we may replace $\sigma^*$ by $\tilde \sigma$,
which has the form
$s_1 I_{n_1} \oplus \cdots \oplus s_k I_{n_k}$.

Now assume that $\sigma^*$ has the form
$s_1 I_{n_1} \oplus \cdots \oplus s_k I_{n_k}$.
Suppose $\rho_i=\rho_j$; say, without loss of generality, $\rho_1 = \rho_2$.
Let $s = (s_1 + s_2)/2$,
$Q = R_2 \otimes I_{n_1} \oplus I_{n-2n_1}$,
and
$\tilde \sigma = (\sigma^* + Q^\dag \sigma^* Q)/2 =
sI_{2n_1} \oplus s_3 I_{n_3} \oplus
\cdots \oplus s_k I_{n_k}$.
Then
\begin{eqnarray*}
\|\rho-\sigma^*\|
&\le& \|\rho-\tilde \sigma\|
= \frac{1}{2} \|(\rho - \sigma^*) + Q^\dag(\rho-\sigma^*)Q\|\\
&\leq& \frac{1}{2} (\|\rho - \sigma^*\| + \|Q^\dag(\rho-\sigma^*)Q\|)
= \|\rho-\sigma^*\|.
\end{eqnarray*}
So, we may replace $\sigma^*$ by $\tilde \sigma$.
\qed

\medskip
\bf Proof of Theorem \ref{2.1}. \rm  Suppose $C$ is a coherence measure induced by a USI norm
$\|\cdot\|$. Then $\|\cdot\|$ satisfies the hypothesis of Proposition \ref{2.3}
and $C$ satisfies Proposition \ref{2.2} (a).
For each natural number $n$, let
$S_n \subseteq \IR^{1\times n}$ be a compact set
satisfying condition (N1)and (N2) for Hermitian matrices
in $M_n$. Given $\rho \in \cD_n$
we have
$$C(\rho) = \min \left\{\|\rho-\sigma\|: \sigma \in \cI_n\right\}.$$
Note the norm computation involves only trace zero matrices.
We may replace every vector $\bc = (c_1, \dots, c_n)\in S_n$
by $\tilde \bc = (c_1-\gamma, \dots, c_n-\gamma)$
so that the largest and smallest entries
of $\tilde \bc$  have the form $c$ and $-c$.
Now, for a trace zero Hermitian matrix $A$,
$$\sum_{j=1}^n (c_j-\gamma) \lambda_j(A)
= \sum_{j=1}^n c_j \lambda_j(A) - \gamma \sum_{j=1}^n \lambda_j(A)
= \sum_{j=1}^n c_j \lambda_j(A) - \gamma \tr A
= \sum_{j=1}^n c_j \lambda_j(A).$$
As a result, the computation of $C(\rho)$ will not be affected.
So, we will assume that the largest and smallest entries
of every vector $\bc \in S_n$ have the form $c$
and $-c$.
Furthermore, we may
replace $C$ by $\alpha C$ for some $\alpha > 0$ and assume that $C(J_2/2) = \|(J_2-I_2)/2\| = 1$.

\medskip\noindent
{\bf Assertion 1.}
\it Let $\bd \in S_4$.
There is a permutation matrix $P \in S_4$
such that $\bd P = (d, d_2, d_3, -d) \in S_4$
with $1 \ge d \ge d_2 \ge d_3 \ge -d \ge -1$.
Moreover, $(1,1,-1,-1) \in S_4$.
Consequently, for any
trace zero Hermitian matrix
$A \in M_4$ with  $\lambda_1 (A) \ge
\lambda_2(A) \ge 0 \ge \lambda_3(A) \ge \lambda_4(A)$,
$$\|A\| =
|\lambda_1(A)| + |\lambda_2(A)| + |\lambda_3(A)| +|\lambda_4(A)|.$$
Furthermore,
$4/3 = C(J_3/3) = C(J_3/3 \oplus [0]).$

\smallskip
\noindent
Proof of Assertion 1. \rm
Let $\bd \in S_4$. By (N1), there is a permutation matrix
$P \in S_4$ such that
$\bd P = (d, d_2, d_3, -d)\in S_4$ such that
$d \ge d_2 \ge d_3 \ge -d$.
Suppose $d > 1$. We may further assume that $d_2 + d_3 \le 0$;
otherwise, replace $\bd$ by the vector
$(d, -d_3, -d_2, -d) \in S_4$.
By (C3) and Proposition \ref{2.3},
if $\rho = J_2/2 \oplus 0_2\in \cD_4$,
then
$$1=  C(J_2/2) = C(\rho)
= \|\rho - \sigma\|$$
for some  $\sigma = \diag(s,s,(1/2-s),(1/2-s)) \in \cI_4$
with $s \in [0, 1/2]$.
The matrix $\rho-\sigma$ has  eigenvalues
$1-s, s-1/2, s- 1/2, -s$.
By (N2),
$$\|\rho-\sigma\| \ge d (1-s)+(d_2+d_3)(s-1/2) -d(-s) \ge d > 1,$$
which is a contradiction.

Next, we prove that $(1,1,-1,-1) \in S_4$.
By (C3) and Proposition  \ref{2.3},
$$
1 = \frac{1}{2}[C(J_2/2) + C(J_2/2)]
= C(J_2/4 \oplus J_2/4) =
\frac{1}{4} \|(J_2-I_2) \oplus (J_2-I_2)\|.
$$
Note that $\frac{1}{4}[(J_2-I_2) \oplus (J_2-I_2)]$ has eigenvalues $1/4,1/4,-1/4,-1/4$.
So, there is a vector in $S_4$ of the form
$(c,c_2,c_3,-c)$ with $1\ge c \ge c_2 \ge c_3 \ge -c\ge -1$ such that
$$
1 = \frac{1}{4}\|(J_2-I_2) \oplus (J_2-I_3)\| =
\frac{1}{4}(2c + c_2-c_3) 
\le \frac{1}{4}(2c+|c_2|+|c_3|) \le c
\le 1.$$
Thus, $c = 1$ and
$(c,c_2,c_3, -c) = (1,1,-1,-1)$.

\medskip
Now, for every vector $\bd \in S_4$, there is a permutation matrix
$P \in S_4$ such that  $\bd P = (d, d_2,d_3, -d)$
with $1 \ge d \ge d_2 \ge d_3 \ge -d \ge -1$
and $(1,1,-1,-1) \in S_4$. By (N2),
for any  trace zero Hermitian matrix
$A \in M_4$ with  $\lambda_1 (A) \ge \lambda_2(A) \ge 0 \ge \lambda_3(A)
\ge \lambda_4(A)$, we have
$$
\|A\| =
\max\{ d \lambda_1(A) + d_2 \lambda_2(A) + d_3 \lambda_3(A)-d\lambda_4(A): (d,d_2, d_3, -d) \in S_4\}
$$
will be attained at the vector $(1,1,-1,-1)$
and $\|A\| = \sum_{j=1}^4
|\lambda_j(A)|.$

\medskip
Now, consider $\rho = J_3/3 \oplus [0] \in \cD_4$.
By Proposition \ref{2.3}, we see that there is $\sigma =
\diag(s,s,s,1-3s) \in \cI_4$ with $s \in [0, 1/3]$ such that
$C(\rho) = \|\rho-\sigma\|.$
Now, $\rho-\sigma$ has eigenvalues
$1-s,1-3s, -s,-s$. Thus,
$$\|\rho-\sigma\| = (1-s) + (1-3s) + s + s = 2-2s$$
which is minimized when $s = 1/3$. By (C3),
$$C(J_3/3) = C(J_3/3 \oplus [0]) = 4/3.$$
The proof of Assertion 1 is complete.

\medskip\noindent
{\bf Assertion 2.}
\it
Let  $\bd \in S_6$. There is a permutation matrix
$P \in M_6$ such that $\bd P =
(d,d_2, \dots, d_5, -d)$ with $1 \ge d \ge d_2
\ge \cdots d_5 \ge -d \ge -1$.
Moreover,
$(1,1,1,-1,-1,-1) \in S_6$.
Consequently, for any
trace zero Hermitian matrix
$A \in M_6$ with  $\lambda_1 (A) \ge \lambda_2(A)
\ge \lambda_3(A) \ge 0 \ge \lambda_4(A)
\ge \lambda_5(A) \ge \lambda_6(A)$,
$$\|A\| =  \sum_{j=1}^6 |\lambda_j(A)|.$$

\noindent
Proof of Assertion 2. \rm
Let $\bd \in S_6$. By (N1),
there is a permutation matrix $P$ such that
$\bd P$ has the form
$(d, d_2, \dots, d_5, -d) \in
S_6$ with $d\ge d_2 \cdots \ge d_5 \ge -d$.
Suppose $d > 1$.
We may assume that $d_2+d_3+d_4+d_5 \le 0$; otherwise, consider
$(d, -d_5, \dots, -d_2, -d) \in S_6$ instead.
By Proposition \ref{2.3}, if
$\rho = J_2/2 \oplus 0_4$, then there is
$\sigma = \diag(2s,2s,1/4-s, 1/4-s, 1/4 -s, 1/4-s) \in \cI_6$
with $s \in [0,1/4]$ such that
$1 = C(\rho) = \|\rho-\sigma\|.$
Now, $\rho-\sigma$ has eigenvalues
$1-2s, s-1/4, s-1/4, s-1/4, s-1/4, -2s$.
By (N2),
$$\|\rho-\sigma\|
\ge d(1-2s) + (s-1/4)(d_2 + \cdots + d_5) + 2ds  \ge d > 1,$$
which is a contradiction.

Next, we show that $(1,1,1,-1,-1,-1) \in S_6$.
Let $\rho = (J_2 \oplus J_2 \oplus J_2)/6$.
By (C3) and Proposition \ref{2.3},
$$1 = \frac{1}{3} C(J_2/2) + \frac{2}{3} C(J_2/4\oplus J_2/4) =
C(\rho) = \|\rho - \rho_\diag\|.$$
Note that $\rho-\rho_\diag$ has three eigenvalues equal to $1/6$ and
three eigenvalues equal to $-1/6$.
Thus, there is $(c, c_2, \dots, c_5, -c) \in S_6$
such that $1 \ge c \ge c_2 \ge \cdots \ge c_5 \ge -c \ge -1$
and
\begin{eqnarray*}
1 &=& \|\rho-\rho_\diag\| = \frac{1}{6}
[2c + (c_2+c_3-c_4-c_5)]\\
& \le&
\frac{1}{6}(2c+|c_2| + |c_3|+|c_4|+c_5|)
\le c\le 1.\end{eqnarray*}
Thus, $c = 1$, and $c_2 = c_3 = 1 = -c_4 = -c_5$.

The proof of the last statement is similar to that in the proof of Assertion 1. The proof of Assertion 2 is complete.

\medskip
To finish the proof of the theorem, consider $\rho = J_2/4 \oplus  J_3/6 \oplus [0]$.
Let  $\sigma = I_2/2 \oplus 0_4$.
Then $\rho-\sigma$ has eigenvalues $1/2, 0, 0, 0, 0 -1/2$,
where the three largest eigenvalues are nonnegative,
and the rest are nonpositive. So,
$\|\rho-\sigma\|= \sum_{j=1}^6 |\lambda_j(\rho-\sigma)| =1$
by Assertion 2.
But then by (C3) and Assertion 1,
$$1 = \|\rho-\sigma\| \ge C(\rho) =
\frac{1}{2}(C(J_2/2) + C(J_3/3 \oplus [0]))
= 1/2(1+4/3) = 7/6,
$$
which is absurd.
\qed

\section{3. Coherence measures associated with the $\ell_{q,p}$-norm}

Recall that, for $1 \leq p, q \leq \infty$, the $\ell_{q,p}$-norm of a matrix $A \in M_n$, with columns $A_1, \dots, A_n$, is the $\ell_q$-norm of the vector formed by the $\ell_p$-norms of the columns of $A$; that is,
$$\ell_{q,p}(A) = \left( \sum_{j=1}^n \ell_p(A_j)^q \right)^{1/q}.$$
By Proposition \ref{2.2} (b), for $\rho \in \cD_n$, we have
 \begin{equation} \label{Cqp}
 C_{q,p}(\rho) = \min\{\ell_{q,p}(\rho-\sigma): \sigma \in \cI_n\} = \ell_{q,p}(\rho - \rho_{\diag}),
 \end{equation}
so these are all easily computed functions of the off-diagonal entries

\begin{thm}\label{T:lqp} The function
$C_{q,p}$ in $(\ref{Cqp})$
is a coherence measure if and only if $q=1$ and $p \in [1,2]$.
\end{thm}

We first establish the necessity  of Theorem \ref{T:lqp}.

\begin{lemma}\label{L:necessity}
If $C_{q,p}$ is a coherence measure then $q=1$ and $p \in [1,2]$.
\end{lemma}

\noindent
\it Proof. \rm
Let $0_n$ be the $n \times n$ zero matrix and let $J_n$ be the $n \times n$ all ones matrix.  Let
$$A = \frac{1}{4} (J_2 \oplus J_2) = \frac{1}{2} \left(\frac{1}{2} J_2 \oplus 0_2 \right) + \frac{1}{2} \left(0_2 \oplus \frac{1}{2} J_2 \right).$$  Then $C_{q,p}(A) = \frac{1}{4}4^{1/q}$ while $\frac{1}{2} C( \frac{1}{2} J_2) + \frac{1}{2} C( \frac{1}{2} J_2) = \frac{1}{2}2^{1/q}$; if $C_{q,p}$ were a coherence measure, then by property (C3) these two quantities must be equal, whence $q=1$.

Now, let $K_1 = (\sin \theta) I_{n} \oplus [0], K_2 = (\cos \theta) I_{n} \oplus [1] \in M_{n+1}$, with $\theta \in [0,\pi/2]$, so $\Lambda(X) = \sum_{j=1}^2 K_j X K_j^{\dag}$ is an incoherent operation.  
Let $A = J_{n+1}$ and $\rho = A/(n+1)$.  Note 
$C_{1,p}(A) = (n+1)C_{1,p}(\rho)$ and 
$C_{1,p}(K_jAK_j^\dag) = (n+1)C_{1,p}(K_j \rho K_j^\dag)$
for $j = 1,2$, and $p \in [1, \infty]$.
Write $c=\cos \theta$, $s=\sin \theta$.
Then
$C_{1,\infty}(A) = n+1, \ C_{1, \infty}(K_1 A K_1^{\dag}) = ns^2, \
C_{1,\infty}(K_2 A K_2^{\dag}) =(n+1)c,$
so
$$
 \sum_{j=1}^2 C_{1,\infty}(K_j A K_j^{\dag}) - C_{1,\infty}(A)
 = n (c-c^2) + c-1 = (nc-1)(1-c)
$$
is positive for $\theta \in (0, \pi/2)$ and $n$ sufficiently large.  This violates property (B3), so $C_{1,\infty}$ is not a coherence measure.

For $p \ne \infty$, we have
\begin{eqnarray*}
C_{1,p}(A) &= & (n+1)n^{1/p} \\
C_{1,p}(K_1 A K_1^\dag) &=& (\sin^2 \theta) n (n-1)^{1/p} \\
C_{1,p}(K_2 A K_2^\dag) &= &(\cos \theta) n^{1/p} + n[(n-1) \cos^{2p} \theta + \cos^p \theta]^{1/p}.
\end{eqnarray*}
Let $f(n,\theta) = C_{1,p}(K_1 A K_1^\dag) + C_{1,p}(K_2 A K_2^\dag ) - C_{1,p}(A)$.  
Then 
\begin{eqnarray*}
f(n,\theta) &=& n^{1/p} \left( c-1 + n(s^2 (1-1/n)^{1/p}
\right.\\
&& \hskip .5in\left. + [(1-1/n)c^{2p} + c^p/n]^{1/p} - 1)\right) \\
&=& n^{1/p} ( c-1 + g(t,\theta)),
\end{eqnarray*}
if $t=1/n$ and
$$g(t,\theta) =  \frac{s^2 (1-t)^{1/p} + [(1-t)c^{2p} + t c^p]^{1/p} - 1}{t}.$$
By l'H\^{o}pital's rule,
\begin{eqnarray*}
\lim_{t \to 0+} g(t,\theta)
&=& \lim_{t \to 0+} \frac{1}{p}\left( -s^2 (1-t)^{1/p - 1}
+ [(1-t)c^{2p}+tc^p]^{1/p - 1} (c^p - c^{2p}) \right) \\
&=& \frac{1}{p} (-s^2 + c^{2-2p}(c^p-c^{2p})) = \frac{1}{p}(c^{2-p} - 1).
\end{eqnarray*}
For $p>2$ we can make this limit arbitrarily large by making $\cos \theta$ sufficiently small.  It follows that $f(n,\theta) > 0$ for $n$ sufficiently large and $\theta$ sufficiently close to $\pi/2$, violating property (B3).
\qed 

The proof of sufficiency for Theorem \ref{T:lqp} is more complicated; the key is to show a norm inequality that may be of independent interest.  To this end, the following notation will be useful.  Let $\IR_+$ be the set of nonnegative real numbers.  Given a set $\Omega$ we shall write $|\Omega|$ for the number of elements in $\Omega$.  Let $\Omega_n = \{1, \dots, n\}$.  Given a subset $\tau \subseteq \Omega_n$, let $\tau^c = \Omega_n \setminus \tau$.  Given a vector $v = (v_1, \dots, v_n) \in \IC^n$ and a nonempty subset $\sigma$ of $\Omega_n$, 
let $v_{\sigma} \in \IC^{|\sigma|}$ be the vector whose entries are $\{v_j : j \in \sigma\}$, ordered by increasing index.

We shall need two technical results first.

\begin{lemma}\label{L:perm}
Let $1 \leq p \leq 2$ and let $\Omega$ be a collection of nonempty subsets of $\Omega_n$.  Suppose $\Omega$ is a cover of $\Omega_n$ and $v \in \IR_+^n$ has no zero entries.  Then
\begin{equation}\label{E:perm}
\left(\max_{\sigma \in \Omega} \ell_p(v_{\sigma}) \right)^{p-2} \ell_2(v)^2 \leq \ell_p(v)^p.
\end{equation}
\end{lemma}

\noindent
{\it Proof.} 
We shall use induction on $|\Omega|$.  When $|\Omega| = 1$, we must have $\Omega = \{ \Omega_n \}$ and the left-hand side of \eqref{E:perm} becomes $\ell_p(v)^{p-2} \ell_2(v)^2 \leq \ell_p(v)^{p-2} \ell_p(v)^2 = \ell_p(v)^p$ as desired.  Now suppose the assertion holds whenever $|\Omega|$ is less than $m$.

Let $\Omega$ be a cover of $\Omega_n$ with $|\Omega| = m$.  Choose $\tau \in \Omega$ so that $\ell_p(v_{\tau}) \geq \ell_p(v_{\sigma})$ for all $\sigma \in \Omega$; if $\tau = \Omega_n$ we are done, so we may assume $\tau \ne \Omega_n$ and write $K = \ell_p(v_{\tau})$.
Let $\tilde{\Omega} = \{\sigma \cap \tau^c: \sigma \in \Omega, \sigma \cap \tau^c \ne \emptyset\}$.  Because $\Omega$ is a cover for $\Omega_n$, $\tilde{\Omega}$ is a cover for $\tau^c$ with $|\tilde{\Omega}| < m$.  Then
\begin{eqnarray*}
K^{p-2} \ell_2(v)^2 
&=&
K^{p-2} \ell_2(v_{\tau})^2 + K^{p-2} \ell_2 (v_{\tau^c})^2 \\
&\leq &
K^{p-2} \ell_p (v_{\tau})^2 + \left( \max_{\mu \in \tilde{\Omega}} \ell_p(v_{\mu}) \right)^{p-2} \ell_2 (v_{\tau^c})^2  \\
&& \hskip .9in
\text{ since } p \leq 2 \text{ and  for all }  \sigma \in \Omega, \ 
\ell_p(v_{\sigma \cap \tau^c}) \leq \ell_p(v_{\sigma}) \leq \ell_p (v_{\tau})\\
&=& \ell_p(v_{\tau})^p +  \left( \max_{\mu \in \tilde{\Omega}} \ell_p(v_{\mu}) \right)^{p-2} \ell_2 (v_{\tau^c})^2 \\
&\leq& \ell_p(v_{\tau})^p + \ell_p(v_{\tau^c})^p
\hskip 1.7in \text{by the induction hypothesis} \\
&=& \ell_p(v)^p
\end{eqnarray*}
as desired.
\qed 

\begin{lemma}\label{L:Lagrange}
Fix $p \in [1,2]$ and let $n \in \IN$.  Let $\Omega$ be a collection of nonempty subsets covering $\{1, \dots, n\}$ and let $v \in \IR_+^n$ be a nonzero vector.  For each $\sigma \in \Omega$ let $b_{\sigma}$ be a nonnegative number.  Suppose
$$\sum_{\sigma \in \Omega} b_{\sigma} = \ell_2(v)^2 \ \
\hbox{ and } \
\sum_{\sigma \subseteq \tau, \sigma \in \Omega} b_{\sigma}
\leq \ell_2 (v_{\tau})^2 \
\hbox{ for } \tau \subseteq \Omega_n.$$
Then
\begin{equation}\label{convex combo}
\sum_{\sigma \in \Omega, v_{\sigma} \ne 0} \ell_p (v_{\sigma})^{p-2} b_{\sigma} \leq \ell_p(v)^p.
\end{equation}
\end{lemma}
\bigskip

\noindent
\it Proof. \rm
We prove this by induction on $n$.  When $n=1$ we must have $\Omega = \{ \{1\} \}$, $v > 0$, and $b_{\{1\}} = v^2$; the assertion clearly holds.

Now suppose \eqref{convex combo} holds whenever the length of $v$ is less than $n$.  Consider the function $f: \IR_+^n \times \IR_+^{|\Omega|}$ defined by $f(v;b) = 0$ if $v_{\sigma} = 0$ for all $\sigma \in \Omega$, and otherwise
\begin{equation*}
f(v;b) = \sum_{\sigma \in \Omega, v_{\sigma} \ne 0}
\ell_p (v_{\sigma})^{p-2} b_{\sigma}.
\end{equation*}
Let $K \subset \IR_+^n \times \IR_+^{|\Omega|}$ be the compact set defined by
\begin{eqnarray*}
K &=&
\{(v;b) \in  \IR_+^n \times \IR_+^{|\Omega|} : \ell_p(v)^p = M >0, 
 \sum_{\sigma \subseteq \tau, \sigma \in \Omega} b_{\sigma} \leq \ell_2 (v_{\tau})^2 \; \forall \tau \subseteq \Omega_n, \\
 && \hskip 3in \text{ with equality when } \tau = \Omega_n \}.
\end{eqnarray*}
In particular, when $(v; b) \in K$ we have $b_{\sigma} \leq \ell_2 (v_{\sigma})^2$, so when $v_{\sigma}$ is nonzero,
$$\ell_p(v_{\sigma})^{p-2} b_{\sigma} \leq \ell_p(v_{\sigma})^p,$$
which approaches zero when $v_{\sigma}$ approaches zero.  Thus $f$ is continuous on $K$ and attains an absolute maximum on $K$; it suffices to show that this maximum does not exceed $M$.
\bigskip

\noindent
Case 1: Suppose the maximum of $f$ is attained on the relative boundary of $K$.  There are three possibilities.
\medskip

\noindent
Subcase (i):  The maximum occurs at some $v \in \IR_+^n$ with a zero entry, say, $v_j = 0$.  We can replace $v \in \IR_+^n$ with $v_{\{j\}^c} \in \IR_+^{n-1}$, $\sigma \in \Omega$ with $\sigma \cap \{j\}^c$, and the result follows by induction.
\medskip

\noindent
Subcase (ii): The maximum occurs when $b_{\sigma} = 0$ for some $\sigma \in \Omega$.  We may replace $\Omega$ by $\Omega \setminus \{\sigma\}$ and use induction on $|\Omega|$; note that when $\Omega$ consists of a single element $\sigma$,
$$f(v;b) \leq \ell_p(v_{\sigma})^{p-2} \ell_2(v_{\sigma})^2 \leq \ell_p(v_{\sigma})^{p-2} \ell_p(v_{\sigma})^2 \leq \ell_p(v)^p.$$
\medskip

\noindent
Subcase (iii): The maximum occurs when $\sum_{\sigma \subseteq \tau, \sigma \in \Omega} b_{\sigma} = \ell_2 (v_{\tau})^2 >0$ for some $\tau \subsetneq \Omega_n$.  
Let
$$\tilde{\Omega} = \{ \sigma \cap \tau^c : \sigma \in \Omega, \sigma \cap \tau^c \ne \emptyset \}$$
and for $\mu \in \tilde{\Omega}$, set
$$\tilde{b}_{\mu} = \sum_{\sigma \cap \tau^c = \mu} b_{\sigma}.$$
Because
$$\ell_2(v)^2 = \sum_{\sigma \in \Omega} b_{\sigma} = \sum_{\sigma \subseteq \tau} b_{\sigma} + \sum_{\sigma \cap \tau^c \ne \emptyset} b_{\sigma} = \ell_2(v_{\tau})^2 + \sum_{\mu \in \tilde{\Omega}} \tilde{b}_{\mu},$$
we have $\sum_{\mu \in \tilde{\Omega}} \tilde{b}_{\mu} = \ell_2(v_{\tau^c})^2$.  Moreover, for all $\nu \subseteq \tau^c$ we have

\begin{eqnarray*}
 \sum_{\mu \subseteq \nu} \tilde{b}_{\mu}
& = &\sum_{\mu \subseteq \nu} \sum_{\sigma \cap \tau^c \mu} b_{\sigma}
=  \sum_{\sigma \subseteq \tau \cup \nu} b_{\sigma}
- \sum_{\sigma \subseteq \tau} b_{\sigma}
\\
&\leq & \ell_2(v_{\tau \cup \nu})^2 - \ell_2 (v_{\tau})^2
= \ell_2(v_{\nu})^2.
\end{eqnarray*}

Then
\begin{eqnarray*}
f(v,b)
& = &\sum_{\sigma \in \Omega, \sigma \subseteq \tau} \ell_p(v_{\sigma})^{p-2} b_{\sigma} + \sum_{\sigma \in \Omega, \sigma \cap \tau^c \ne \emptyset} \ell_p(v_{\sigma})^{p-2} b_{\sigma} \\
&\leq& \sum_{\sigma \in \Omega, \sigma \subseteq \tau} \ell_p(v_{\sigma})^{p-2} b_{\sigma} + \sum_{\sigma \in \Omega, \sigma \cap \tau^c \ne \emptyset} \ell_p(v_{\sigma \cap \tau^c})^{p-2} b_{\sigma} \\
&=& \sum_{\sigma \in \Omega, \sigma \subseteq \tau} \ell_p(v_{\sigma})^{p-2} b_{\sigma} + \sum_{\mu \in \tilde{\Omega}} \ell_p (v_{\mu})^{p-2} \tilde{b}_{\mu}.
\end{eqnarray*}
By induction (since the lengths of $v_{\tau}$ and of $v_{\tau^c}$ are less than $n$), the last two terms do not exceed $\ell_p(v_{\tau})^p$ and $\ell_p(v_{\tau^c})^p$ respectively and the result follows.
\bigskip

\noindent
Case 2: Suppose the maximum of $f$ is attained in the relative interior of $K$.
Using Lagrange multipliers we conclude that $\nabla f = \lambda_1 \nabla g_1 + \lambda_2 \nabla g_2$ where
$$\qquad g_1(v;b) = \ell_p(v)^p - M = 0
\quad \hbox{ and }\quad
g_2(v;b) = \ell_2(v)^2 - \sum_{\sigma \in \Omega} b_{\sigma} = 0.$$
From the partial derivative with respect to $b_{\sigma}$, $\sigma \in \Omega$, we have
\begin{equation*} 
\ell_p(v_{\sigma})^{p-2} = 0 + \lambda_2 (-1),
\end{equation*}
whence $\ell_p(v_{\sigma})$ equals a constant $K$ for all $\sigma \in \Omega$.
Thus the maximum of $f$ is given by
$$f(v;b) = \sum_{\sigma \in \Omega} K^{p-2} b_{\sigma} = K^{p-2} \ell_2(v)^2$$
and the result then follows by Lemma \ref{L:perm}.
\qed

The following result is known, e.g., see \cite{KL}.
We include a proof for completeness.

\begin{lemma}\label{L:extreme points}
If $B$ is an extreme point for the unit ball for the $\ell_{1,p}$-norm, then $B$ has exactly one nonzero column.
\end{lemma}

\noindent
\it Proof. \rm
Clearly $B \ne 0$.  We prove the contrapositive.  Suppose $\ell_{1,p}(B) = 1$ and $B$ has more than one nonzero column; without loss of generality, we may suppose that the first two columns are nonzero.  Let $b_j$ be the $j$th column of $B$.  Let $\epsilon = \frac{1}{2} \min \{\ell_p(b_1), \ell_p(b_2)\}$.  Then
$$
B = \frac{1}{2}
\begin{bmatrix} \left( 1+ \frac{\epsilon}
{\ell_p(b_1)} \right) b_1 & \left( 1 - \frac{\epsilon}{\ell_p(b_2)} \right) b_2 & b_3 & \dots & b_n \end{bmatrix} 
+ \frac{1}{2} 
\begin{bmatrix} \left( 1 - \frac{\epsilon}{\ell_p(b_1)} \right) b_1 & \left( 1+ \frac{\epsilon}{\ell_p(b_2)} \right) b_2 & b_3 & \dots & b_n 
\end{bmatrix}$$
is the average of two distinct matrices with norm 1, so $B$ is not an extreme point.
\qed

The next result provides the main idea for showing sufficiency in Theorem \ref{T:lqp}.  The seminorm defined is in fact a norm, but that is not needed for our purposes.

\begin{prop}\label{P:contraction}
Let $p \in [1,2]$ and let $\{K_1, \dots, \K_m\}$ be a set of incoherent Kraus operators in $M_{N,n}$.  Define a seminorm $\| \cdot \|$ on $M_n$ by
$$\|A\| = \sum_{k=1}^m \ell_{1,p}(K_k A K_k^{\dag}).$$
Then $\|A\| \leq \ell_{1,p}(A)$ for all $A \in M_n$.
\end{prop}

\noindent
\it Proof. \rm
Let $\cB$ and $\cB_{1,p}$ be the unit balls in $M_n$ for $\| \cdot \|$	and $\ell_{1,p}$ respectively.  Then $\|A\| \leq \ell_{1,p}(A)$ for all $A$ if and only if the unit ball for the $\ell_{1,p}$-norm lies inside the unit ball for the $\| \cdot \|$-seminorm.  By convexity, it suffices to show that each extreme point $B$ of the unit $\ell_{1,p}$-ball has seminorm $\|B\| \leq 1$.

By Lemma \ref{L:extreme points}, such an extreme point $B$ has exactly one nonzero column.  Let $e_j$ be the vector whose only nonzero entry is a $1$ in the $j$th position.  We may write $B = ve_j^{\dag}$ for some $\ell_p$-unit vector $v \in \IC^n$; without loss of generality, we may assume $j=1$.  Thus we must show that
\begin{equation}\label{goal}
\sum_{k=1}^m  \ell_{1,p} (K_k ve_1^\dag  K_k^\dag) \leq 1.
\end{equation}
for all $v \in \IC^n$ with $\ell_{p}(v) = 1$.
For such a $v$, write $v = \sum_{j=1}^n v_j e_j \in \IC^n$. Let
$$F = \begin{bmatrix} K_1 \\ K_2 \\ \vdots \\ K_m \end{bmatrix}, \qquad w = Fv = \begin{bmatrix} w_1 \\ w_2 \\ \vdots \\ w_m \end{bmatrix},$$
where $w_k = K_k v$.  It is important to note that, because $\sum_{k=1}^m K_k^{\dag} K_k = I$, $F$ is an isometry (for $\ell_2$).

Since $K_k \cI_n K_k^{\dag}$ is diagonal, each column of $K_k$ has at most one nonzero entry (see \cite[Theorem 1]{Yao}, or simply note that, if $K_k e_j$ had nonzero entries in the $p$- and $q$-positions, then $K_k e_j e_j^{\dag} K_k^{\dag}$ would have a nonzero entry in the $(p,q)$-position).  Thus we may write $K_k e_j = c_{kj} e_{\sigma_k(j)}$; here $\sigma_k$ is a map from $\{1,\dots, n\}$ to $\{1, \dots, N\}$.  Then
$$\ell_{1,p} (K_k v e_1^\dag  K_k^\dag) = \ell_{1,p} (w_k \bar{c}_{k1} e_{\sigma_k(1)}^\dag) = |c_{k1}| \ell_p(w_k),$$
so
$$
\sum_{k=1}^m \ell_{1,p} (K_k v e_1^{\dag} K_k^{\dag})=
\sum_{k=1}^m |c_{k1}| \ell_p (w_k)
\leq \sqrt{\sum_k |c_{k1}|^2} \sqrt{\sum_k\ell_p(w_k)^2}
$$
by the Cauchy-Schwarz inequality.  Since $F$ is an isometry, $\sum_k |c_{k1}|^2 = \ell_2(Fe_1)^2 = 1$, so \eqref{goal} will hold if
we can prove
\begin{equation}\label{goal2}
\sum_{k=1}^m \ell_p (w_k)^2 \leq 1.
\end{equation}
As an aside, for $p=2$ the proof is complete, for $\sum_k \ell_2 (w_k)^2 = \ell_2(w)^2 = \ell_2(v)^2 = 1$
because $F$ is an isometry.  For $1 \leq p < 2$, however, the proof continues.

Define $\cJ_{k,s} = \{j : \sigma_k(j) = s\}$.  Note that
$$
w_k = 
K_k \left(\sum_{j=1}^n v_j e_j \right) = \sum_{j=1}^n v_j c_{kj} e_{\sigma_k(j)} 
= \sum_s \left(\sum_{j \in \cJ_{k,s}} v_j c_{kj} \right) e_s.
$$
Let $v_{k,s}$  be the vector of length $|\cJ_{k,s}|$ 
with entries equal to $v_j, j \in \cJ_{k,s}$, and let
$w_{k,s} = \sum_{j \in \cJ_{k,s}} v_j c_{kj}$ be the $s$th entry of $w_k$.  
Then 
$$
\ell_p(w_k)^2 = \left( \sum_s |w_{k,s}|^p \right)^{2/p} 
= \left( \sum_{s, v_{k,s} \ne 0}\ell_p(v_{k,s})^p \, 
\frac{|w_{k,s}|^p}{\ell_p(v_{k,s})^p} \right)^{2/p}
\leq   
\sum_{s, v_{k,s} \ne 0} 
\ell_p(v_{k,s})^p \, \frac{|w_{k,s}|^2}{\ell_p(v_{k,s})^2}
$$
since $f(x) = x^{2/p}$ is convex for $p \in [1,2]$ and $\sum_s \ell_p(v_{k,s})^p = \ell_p(v)^p = 1$.

Thus \eqref{goal2} will hold if we can show that
\begin{equation}\label{new goal}
\sum_{k,s, v_{k,s} \ne 0} \ell_p(v_{k,s})^{p-2} \, |w_{k,s}|^2 \leq 1.
\end{equation}
Note that $v_{k,s}$ consists of the entries of $v$ whose indices correspond to nonzero entries in the $s$th row of $K_k$, so we may regroup the sum as follows.  Given a nonempty subset $\sigma \subseteq \{1, \dots, n\}$, recall that $v_{\sigma} \in \IC^{|\sigma|}$ consists of the entries of $v$ whose indices lie in $\sigma$.   Define $$b_{\sigma} = \sum_i |(Fv)_i|^2,$$
where the sum is over all $i$ such that $\{j : F_{ij} \ne 0 \} = \sigma$, and let $\Omega$ be the collection of all nonempty $\sigma$ for which there exists an $i$ such that $\{j : F_{ij} \ne 0 \} = \sigma$.  Then \eqref{new goal} is equivalent to
\begin{equation}\label{most general}
\sum_{\sigma \in \Omega, v_{\sigma} \ne 0} \ell_p(\vec{v}_{\sigma})^{p-2} \, b_{\sigma} \leq 1,
\end{equation}
which follows from Lemma \ref{L:Lagrange} (note that the hypotheses for the lemma are satisfied because $F$ is an isometry).
\qed
\bigskip

\medskip
\bf Proof of Theorem \ref{T:lqp}. \rm
Necessity was shown by Lemma \ref{L:necessity}.  Let $p \in [1,2]$.  By Proposition \ref{2.2}, to show that $C_{1,p}$ is a coherence measure it suffices to show that property (B3) holds.  Let $\{K_1, \dots, K_m\}$ be a set of incoherent Kraus operators.  Let $\rho \in M_n$, $p_j = \tr K_j \rho K_j^{\dag}$, and $\rho_j = \frac{1}{p_j} K_j \rho K_j^{\dag}$.  Then
$$
\sum_{j=1}^m p_j C_{1,p}(\rho_j) =
\sum_{j=1}^m \ell_{1,p} (K_j \rho K_j^{\dag} 
- (K_j \rho K_j^{\dag})_{\diag}) \leq
 \sum_{j=1}^m \ell_{1,p} (K_j (\rho - \rho_{\diag}) K_j^{\dag})
$$
because $K_j \cI_n K_j^{\dag}$ is diagonal for all $j$.  By Proposition \ref{P:contraction}
$$ \sum_{j=1}^m \ell_{1,p} (K_j (\rho - \rho_{\diag}) K_j^{\dag}) \leq \ell_{1,p} (\rho - \rho_{\diag}) = C_{1,p}(\rho),$$
so (B3) holds.
\qed

\section{4. Conclusion and further research}
In this article, we study coherence measures induced by norm functions.
It is shown that
no unitary similarity invariant norm induces a coherence measure; this generalizes the negative result for Schatten $p$-norms.  On the other hand, the $\ell_{q,p}$-norm can induce a coherence measure, but if and only if $q=1$ and $1 \leq p \leq 2$.  This provides a new class of potentially useful coherence measures.
It would be interesting to extend
our techniques to study
quantum coherence in multipartite
systems and related problems; \cite{Napoli,Piani,Streltsov,Yao,RPJT,XLF,XZZ}.

\bigskip

\textbf{ACKNOWLEDGMENTS}

C.K. Li is an affiliate member of the Institute for Quantum Computing, 
University of
Waterloo. His research was supported by the Simons Foundation
Grant 351047. This research of Y. Jing was supported by National 
Natural Science Foundation of China under Grant No.11801123.
\bigskip

\textbf{DATA AVAILABILITY}

Data sharing is not applicable to this article as no new data were created or analyzed in this study.

\end{document}